\documentclass[12pt]{article}
\usepackage{amsmath,amssymb,mathrsfs,url,graphicx,rotating}

\title{Parameter Diagrams of the GRW and CSL Theories of Wave Function Collapse}
   
\author{
William Feldmann\footnote{Department of Physics and Engineering Physics,
	Stevens Institute of Technology, Castle Point on Hudson,
	Hoboken, NJ 07030, USA. 
	E-mail: wfeldman@stevens.edu},
Roderich Tumulka\footnote{Department of Mathematics,
     Rutgers University, Hill Center, 
     110 Frelinghuysen Road, Piscataway, NJ 08854-8019, USA.
     E-mail: tumulka@math.rutgers.edu}
}
\date{November 15, 2011}

\newcommand{\be}{\begin{equation}}
\newcommand{\ee}{\end{equation}}
\newcommand{\RRR}{\mathbb{R}}
\newcommand{\CCC}{\mathbb{C}}

\newcommand{\scp}[2]{\langle #1|#2 \rangle}
\newcommand{\vc}{\boldsymbol{c}}

\newcommand{\vx}{\boldsymbol{x}}
\newcommand{\vy}{\boldsymbol{y}}

\newcommand{\sI}{\mathscr{I}}

\begin{document}
\maketitle
\begin{abstract}
It has been hypothesized that the time evolution of wave functions might include collapses, rather than being governed by the Schr\"odinger equation. The leading models of such an evolution, GRW and CSL, both have two parameters (or new constants of nature), the collapse width $\sigma$ and the collapse rate $\lambda$. We draw a diagram of the $\sigma\lambda$-plane showing the region that is empirically refuted and the region that is philosophically unsatisfactory. 

\medskip

\end{abstract}

\section{Introduction}

\begin{figure}[h]
\begin{center}
\begin{minipage}{70mm}
(a) \includegraphics[scale=0.6]{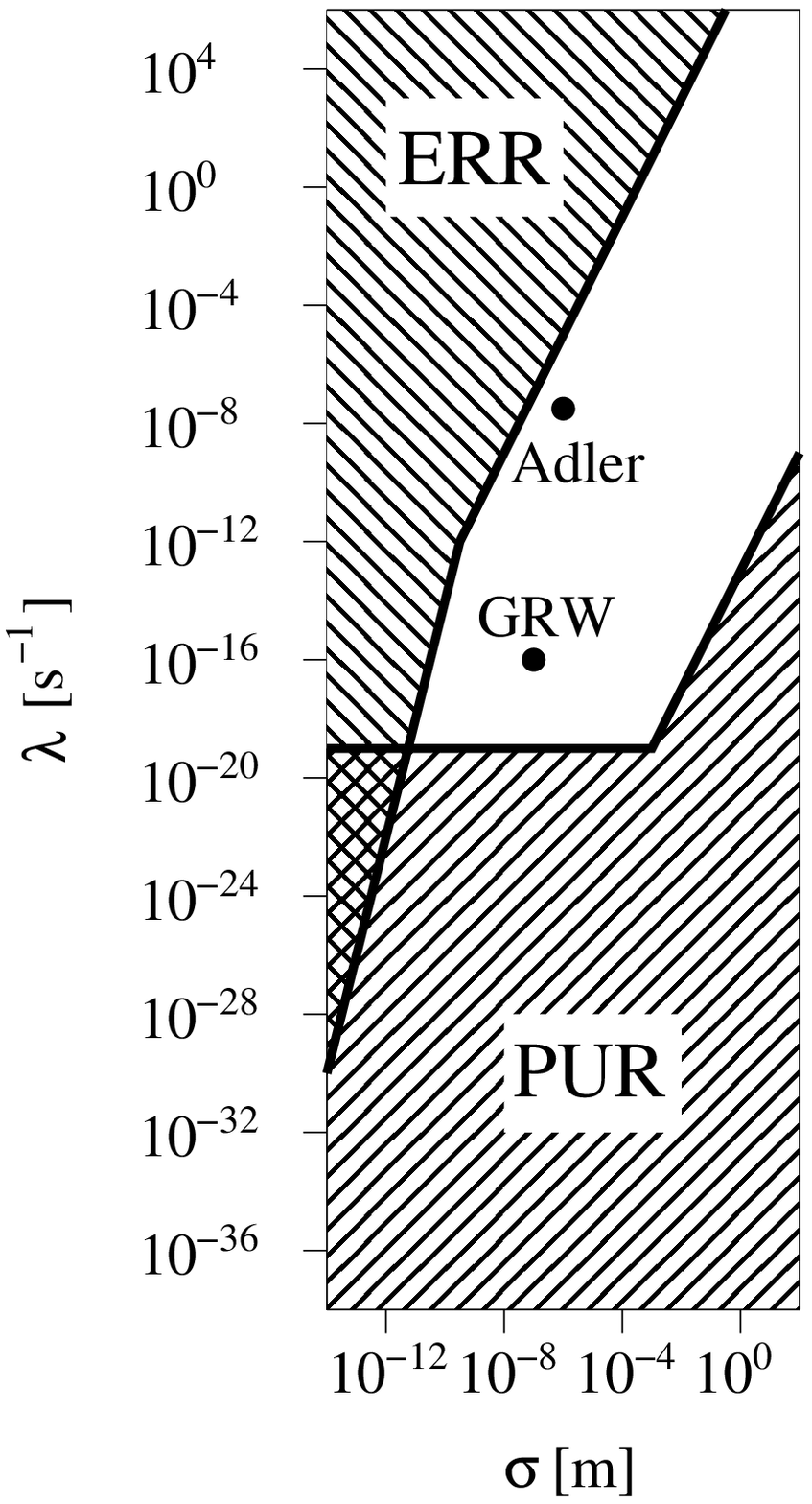}
\end{minipage}
\begin{minipage}{70mm}
(b) \includegraphics[scale=0.6]{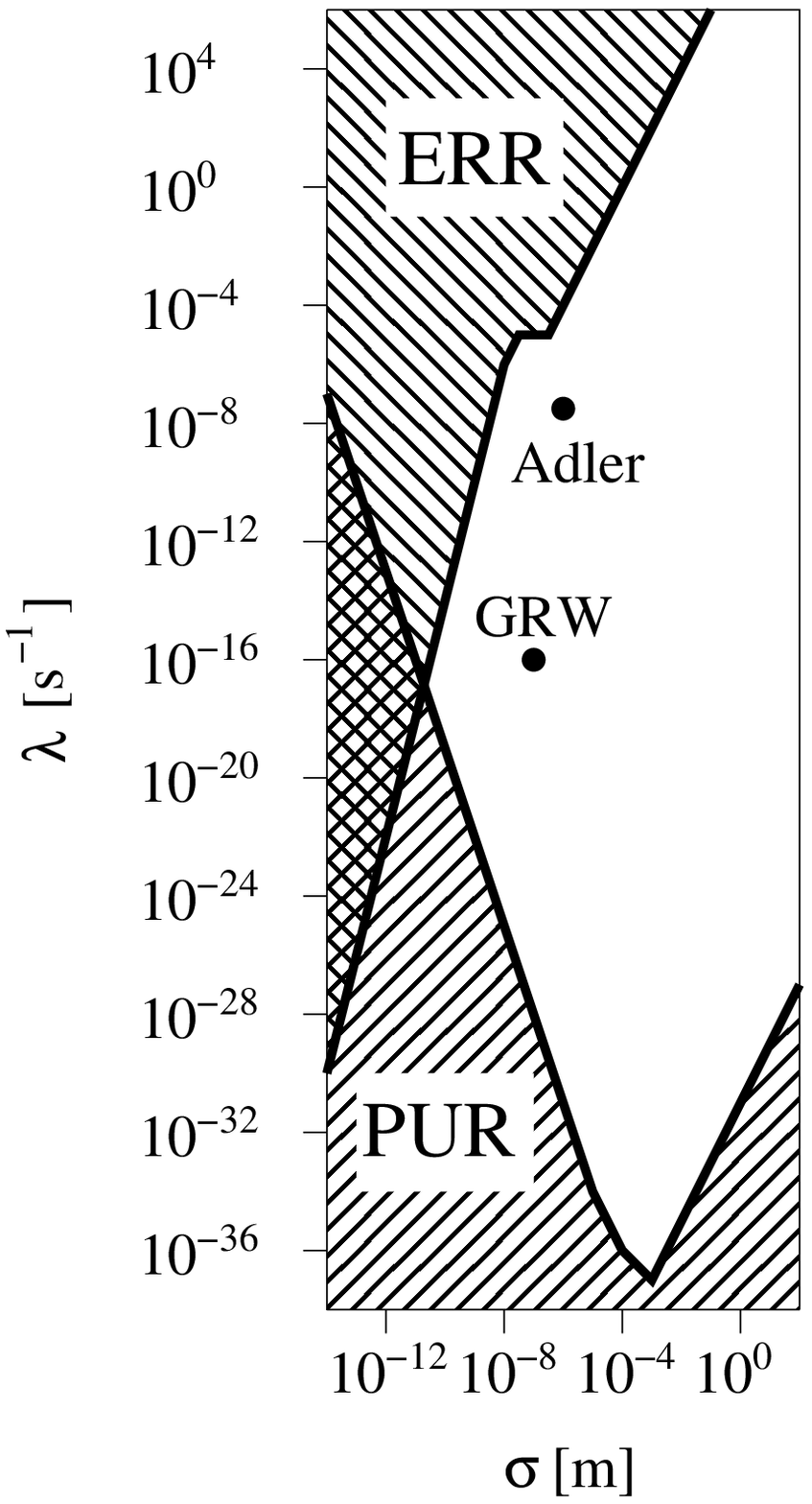}
\end{minipage}
\caption{\label{fig:1}Parameter diagram (log-log-scale) of  (a) the GRW theory, (b) the CSL theory, in both cases with the primitive ontology given by the matter density function. ERR = empirically refuted region, PUR = philosophically unsatisfactory region. GRW's \cite{GRW86} and Adler's \cite{Adl07} choice of parameters are marked.}
\end{center}
\end{figure}

We provide an up-to-date version of the \emph{parameter diagram} (Fig.~\ref{fig:1}) for the GRW \cite{GRW86,Bell87} and CSL \cite{Pe89} theories. These theories solve the conceptual problems of quantum mechanics by postulating a stochastic process replacing the Schr\"odinger equation that will avoid macroscopic superpositions such as Schr\"odinger's cat \cite{Bell87,BG03,Ghi07}. Both GRW and CSL involve two parameters (or new constants of nature), the collapse width $\sigma$ and the collapse rate $\lambda$. A notation often used \cite{GRW86,BG03,Adl07} instead of $\sigma$ is $\alpha=1/(2\sigma^2)$. The parameter diagram, introduced by Collett, Pearle, Avignone and Nussinov \cite{CPAN95,CP03} for CSL, is a diagram of the parameter plane with axes $\sigma$ and $\lambda$. The values of $\sigma$ and $\lambda$ can in principle be measured if our world is governed by GRW or CSL, as the empirical predictions of these theories deviate (slightly) from those of quantum mechanics and depend on $\sigma$ and $\lambda$. At present, only certain combinations $(\sigma,\lambda)$ can be excluded as leading to predictions that disagree with experimental findings; these points form the \emph{empirically refuted region} (ERR) of the parameter plane. Certain values for the parameters, viz., $\sigma=10^{-7}\,\mathrm{m}$ and $\lambda=10^{-16}\,\mathrm{s}^{-1}$, were suggested by Ghirardi, Rimini, and Weber \cite{GRW86}, to which we refer as the ``GRW values''; different values were suggested by Adler \cite{Adl07}, $\sigma=10^{-6}\,\mathrm{m}$ and $\lambda=3\times10^{-8}\,\mathrm{s}^{-1}$. Both choices (as well as the standard quantum mechanical predictions) are compatible with all presently available experimental data. An experimental decision between GRW (or CSL) and standard quantum mechanics (SQM) can come about in two ways: Either the predictions of GRW/CSL get confirmed for a particular pair $(\sigma,\lambda)$ (and SQM gets falsified), or the ERR gets enlarged so much by new data that it covers, together with the \emph{philosophically unsatisfactory region} (PUR), the entire quadrant $\sigma>0$, $\lambda>0$ of the parameter plane (so that GRW/CSL gets falsified). We count those points as ``philosophically unsatisfactory'' for which the GRW/CSL model does not work as intended and fails to produce a picture of macroscopic reality that agrees with what humans normally think macroscopic reality is like, and thus fails to solve the measurement problem of quantum mechanics. The parameter diagram shows the ERR and the PUR. It conveys at a glance information about the empirical restrictions to the GRW/CSL theory. 

In Section \ref{sec:definitions} we give the exact definitions of the GRW and CSL models we use. In Sections \ref{sec:ERR} and \ref{sec:PUR} we describe how we determine the ERR and the PUR, respectively.

\begin{figure}[h]
\begin{center}
\begin{minipage}{70mm}
(a) \includegraphics[scale=0.6]{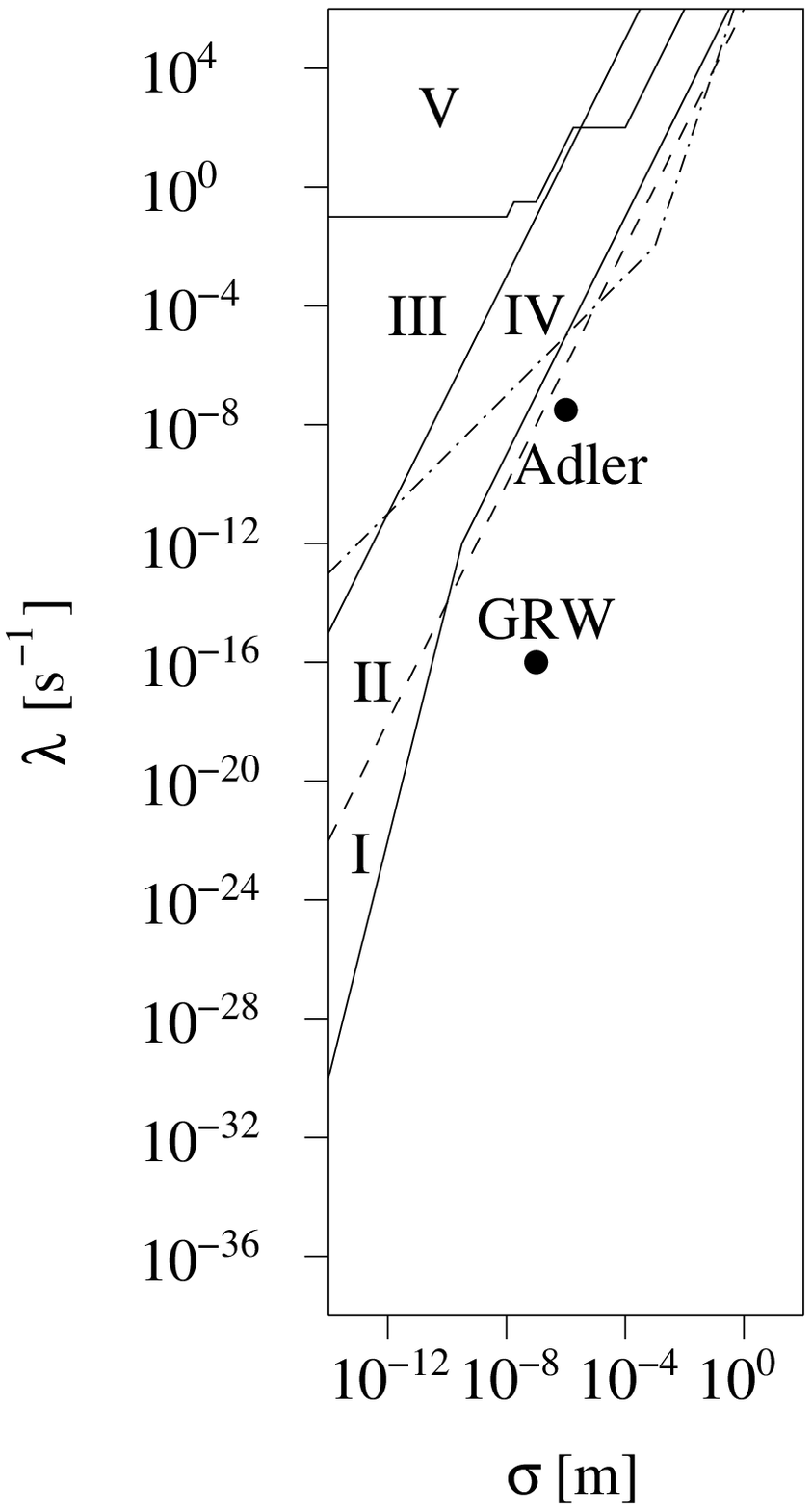}
\end{minipage}
\begin{minipage}{70mm}
(b) \includegraphics[scale=0.6]{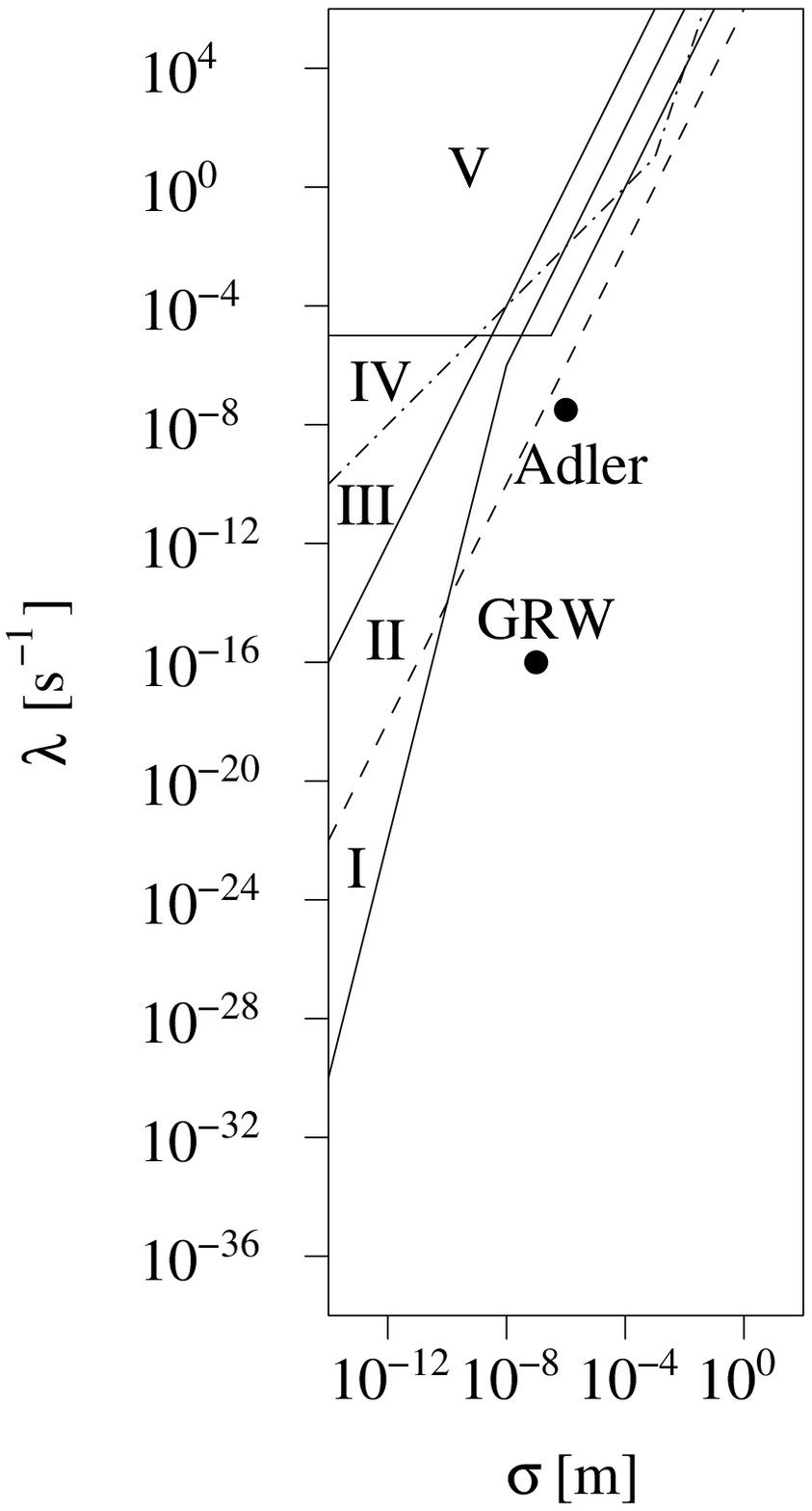}
\end{minipage}
\caption{\label{fig:2}The ERR broken up into regions excluded by different types of observations, (a) for GRW, (b) for CSL: I = spontaneous x-ray emission, II = spontaneous warming of the intergalactic medium (dashed line), III = spontaneous warming of air, IV = decay of supercurrents (dashed-and-dotted line), V = diffraction experiments.  See Section~\ref{sec:ERR} for discussion.}
\end{center}
\end{figure}

\begin{figure}[ht]
\begin{center}
\begin{minipage}{70mm}
(a) \includegraphics[scale=0.6]{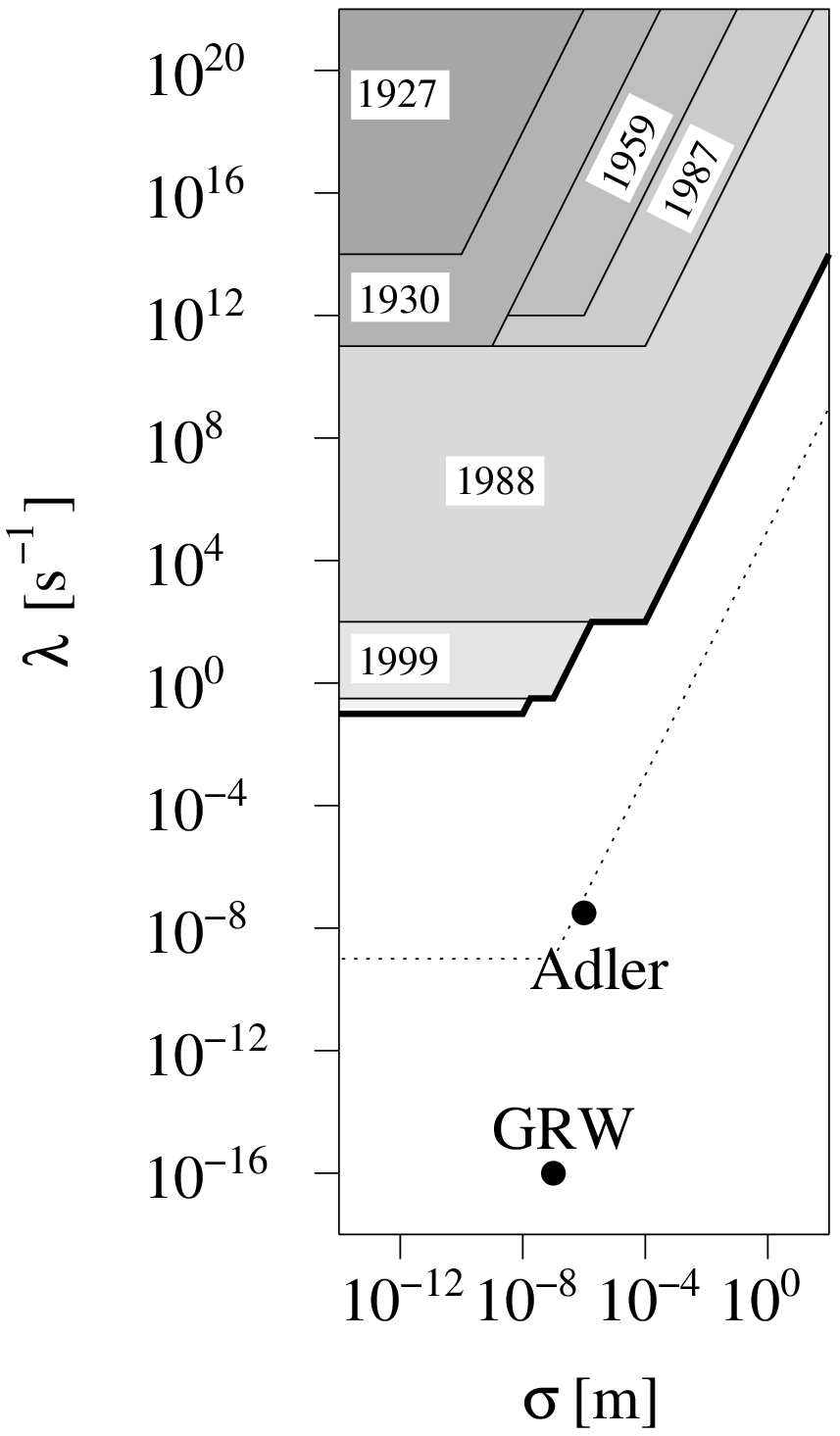}
\end{minipage}
\begin{minipage}{70mm}
(b) \includegraphics[scale=0.6]{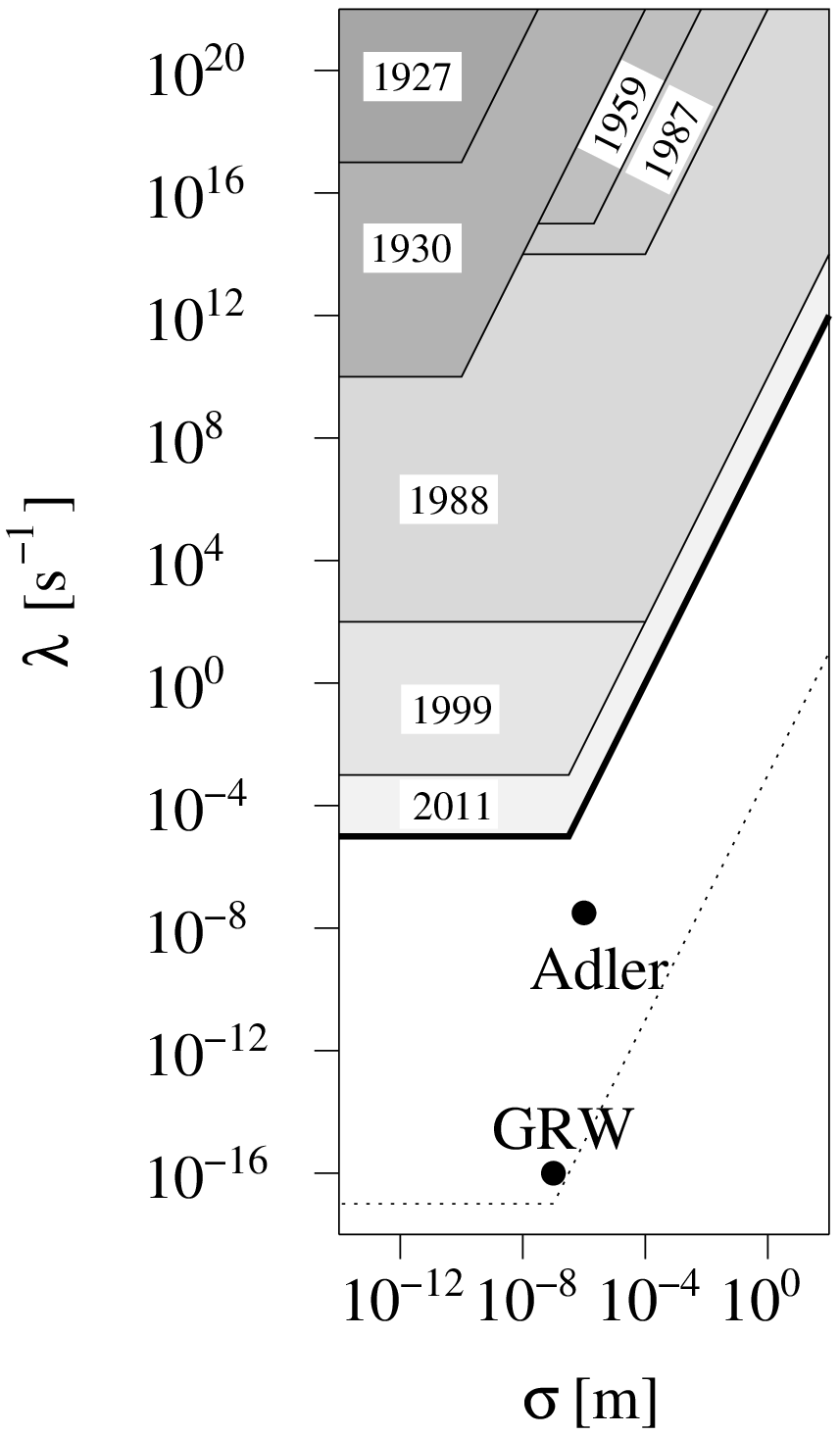}
\end{minipage}
\caption{\label{fig:3}Growth of the subregion V (diffraction experiments) of the ERR as more and better data became available, (a) for GRW, (b) for CSL. Dotted line: proposed future experiments. Note that the interval used on the $\lambda$-axis is different from that of the other figures.}
\end{center}
\end{figure}

We can also identify the subregion of the ERR excluded by each specific type of observation (Fig.~\ref{fig:2}) in order to visualize the strength and relevance of each type of observation. Likewise, we can show how the ERR grows over the years (Fig.~\ref{fig:3}). The parameter diagram also brings out clearly that the empirical rejection of GRW/CSL will require a philosophical decision (viz., where to draw the boundary of the PUR), and that there is an entire region, besides the GRW values and the Adler values, of acceptable values of the parameters. 

The PUR depends on the choice of the variable representing matter in 3-space used in the derivation of predictions; the technical name for this variable is the ``primitive ontology'' (PO). At least two choices of PO have been suggested for the GRW theory: ``flashes'' (GRWf) \cite{Bell87,Tum06a} and the ``matter density function'' (GRWm) \cite{BGG95,Gol98}; see \cite{AGTZ06} for discussion. For the CSL theory, because it does not work with flashes, the PO must be taken to be the matter density function (CSLm). 

\begin{figure}[ht]
\begin{center}
\includegraphics[scale=0.6]{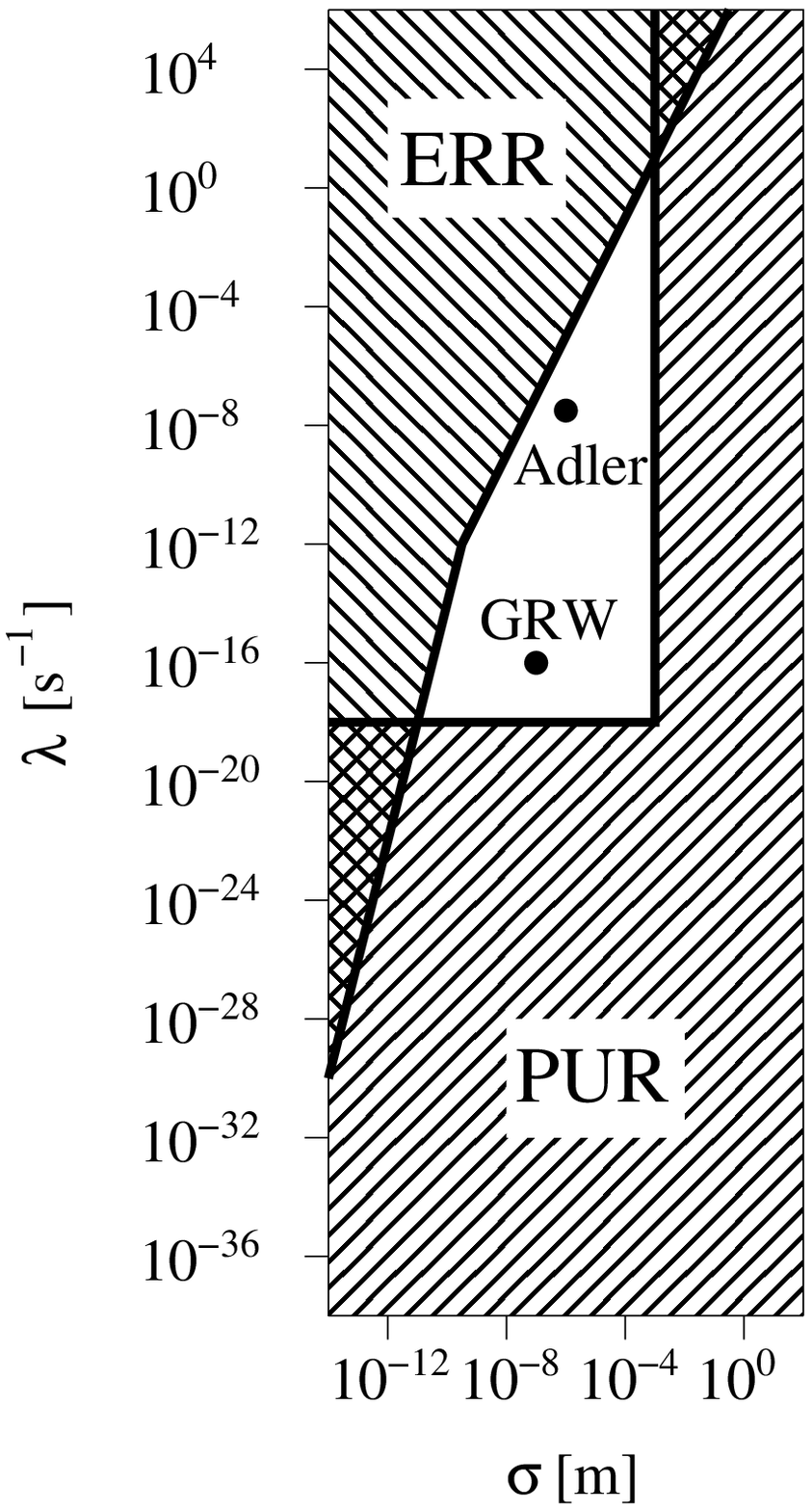}
\caption{\label{fig:GRWf}Parameter diagram of the GRW theory with the primitive ontology given by flashes. The ERR is the same as in Fig.~\ref{fig:1}(a), the PUR is different.}
\end{center}
\end{figure}

The boundaries of the ERR and the PUR as drawn in the Figures are subject to various uncertainties:
\begin{itemize}
\item A careful analysis of how the GRW/CSL dynamics affects the outcomes of a particular experiment or observation requires much research effort. We often have to resort to very rough estimates, substantial idealizations, and gross simplifications. We try to be conservative (i.e., rather draw the ERR too small than too big).  Specific uncertainties will be discussed in Section~\ref{sec:ERR}.
\item A different type of uncertainty concerns the PUR: We have to judge what should be considered philosophically unsatisfactory. Obviously, there is room for disagreement here. Again, we try to be conservative (i.e., rather draw the PUR too small than too big). In particular, Gisin and Percival \cite{GP93} and Adler \cite{Adl07} have expressed the view that collapse should be triggered already by the formation of latent images on nuclear film, rather than by the development of the film; this view corresponds to a much larger PUR than we draw.
\end{itemize}

\section{Definitions Used}
\label{sec:definitions}

\subsection{GRW}

The original version of the GRW model \cite{GRW86,Bell87} was formulated for distinguishable particles; for identical particles, we follow \cite{Tum06b} and the second model in \cite{DS95}. We follow \cite{PS94,BNR95,JPR04} in taking the collapse rate per particle to be proportional to the mass. Thus, the model is as follows. For a system of $N$ ``particles,'' among which $N_k$ belong of species number $k$, namely those with labels $i\in \sI_k\subseteq\{1,\ldots,N\}$, and have mass $m_k$, collapses of type $k$ occur independently of those of other types at rate
\be\label{rateGRW}
\lambda_k = \frac{m_k}{m_p}\lambda\,,
\ee
where $m_p$ is the mass of a proton. If a collapse of type $k$ occurs at time $T$, the wave function $\psi_t:\RRR^{3N}\to\CCC^d$ changes according to
\be\label{collapse}
\psi_{T+}(\vx_1,\ldots,\vx_N) =  
\frac{1}{Z} \Big(\sum_{i\in\sI_k} g(\vc-\vx_i) \Big)^{\! 1/2} \psi_{T-}(\vx_1,\ldots,\vx_N)\,,
\ee
where $g$ is the Gaussian of width $\sigma$,
\be\label{gdef}
g(\vx) = \frac{1}{(2\pi\sigma^2)^{3/2}} e^{-\vx^2/2\sigma^2}\,,
\ee
$Z$ is the normalizing factor,
\be\label{Zdef}
Z = \int_{\RRR^{3N}}d^3\vx_1\cdots d^3\vx_N\, \sum_{i\in\sI_k} g(\vc-\vx_i) 
\big|\psi_{T-}(\vx_1,\ldots,\vx_N)\big|^2 \,,
\ee
and the center $\vc\in\RRR^3$ is chosen randomly with probability density $\rho(\vc) = Z$. Between collapses, the wave function evolves according to the Schr\"odinger equation
\be\label{Schr}
i\hbar\frac{\partial\psi}{\partial t} = H\psi\,.
\ee
Depending on the choice of PO, either a flash occurs at the space-time point $(T,\vc)$ for every collapse, or the matter density function is given by
\be\label{mdef}
m(\vx,t) = \sum_k \sum_{i\in \sI_k} m_k \int_{\RRR^{3N}} d^3\vx_1\cdots d^3\vx_N \, 
\delta^3(\vx-\vx_i)\,
\big|\psi_t(\vx_1,\ldots,\vx_N)\big|^2\,.
\ee

\subsection{CSL}

We follow \cite{PS94,BNR95,JPR04} in taking the collapse rate constant $\lambda_k$ of species $k$ to be
\be\label{rateCSL}
\lambda_k = \Big(\frac{m_k}{m_p}\Big)^2 \lambda\,.
\ee
The wave function $\psi_t:\RRR^{3N}\to\CCC^d$ evolves according to the Ito-type equation
\begin{align}
d|\psi_t\rangle &= \bigg[ -\frac{i}{\hbar} H\, dt + \int_{\RRR^3} d^3\vx \sum_k \sqrt{\lambda_k} 
\Big(N_k(\vx)-\scp{\psi_t}{N_k(\vx)|\psi_t}\Big) \, dB(\vx,t) \nonumber\\
&\quad -\frac{1}{2} \int_{\RRR^3} d^3\vx \bigg(\sum_k\sqrt{\lambda_k}
\Big(N_k(\vx) - \scp{\psi_t}{N_k(\vx)|\psi_t}\Big) \bigg)^{\!2} dt \bigg] |\psi_t\rangle\,,
\label{Ito}
\end{align}
where $B(\vx,t)$ is a family of Wiener processes satisfying
\be
\overline{dB(\vx,t)}=0\mbox{ and }\overline{dB(\vx,t)\,dB(\vy,t)}=\delta^3(\vx-\vy)\, dt\,,
\ee
and the operators $N_k(\vx)$ are defined to be the multiplication operators
\be
N_k(\vx) \, \psi(\vx_1,\ldots,\vx_N) = \sum_{i\in\sI_k} g(\vx-\vx_i)\, \psi(\vx_1,\ldots,\vx_N)
\ee
with $g$ as in Eq.~\eqref{gdef}.

The matter density function is given by Eq.~\eqref{mdef}.

\section{Considerations for Drawing the ERR}
\label{sec:ERR}

As a preliminary consideration, we note the known fact that many mild GRW collapses have an effect similar to few strong collapses. In order to obtain a quantitative version of this relation, we note that for $H=0$ (i.e., pure collapse dynamics), $n$ GRW collapses of width $\sigma$ acting on the $i$-th particle have the same effect as one collapse of width $\sigma/\sqrt{n}$:
\be\label{ncollapses}
e^{-\frac{(\vx_i-\vc_1)^2}{4\sigma^2}} \cdots
e^{-\frac{(\vx_i-\vc_n)^2}{4\sigma^2}} \psi(\vx_1,\ldots,\vx_N) =
C\,e^{-n\frac{(\vx_i-\vc)^2}{4\sigma^2}}
\psi(\vx_1,\ldots,\vx_N)\
\ee
with arbitrary centers $\vc_1,\ldots,\vc_N\in\RRR^3$, $\vc$ their average, and $C>0$ a constant. This observation suggests that the cumulative strength of the collapses per unit time is given by
\be\label{lambdasigma2}
\frac{\lambda}{\sigma^2}\,.
\ee

\subsection{Diffraction Experiments}

A GRW collapse can destroy interference if it hits the interfering particle during its flight between the grating and the detecting screen. Thus, every successful diffraction experiment puts bounds on $\sigma$ and $\lambda$ \cite{Pe84}. Presumably, collapses shortly before the arrival on the screen do not affect the interference pattern much. However, it is not easy to confidently determine the subinterval of the flight time during which collapses disturb the pattern significantly; we would welcome research into estimating that subinterval. For lack of a better estimate, we use for most experiments $1/\tau$, where $\tau$ is the time of flight, as the upper bound on the collapse rate $\lambda_k$ posed by the observation of interference (even though $1/\tau$ may often be too low), with $\lambda_k$ given by \eqref{rateGRW}. We further assume that a single collapse does not spoil the interference if the collapse width $\sigma$ exceeds the distance $d$ between the centers of the slits of the grating. In this case, however, several collapses may have a cumulative effect similar to that the effect of a single collapse of smaller width. For this reason, we assume that the interference pattern is not spoiled if $\lambda_k/\sigma^2< 1/(d^2\tau)$. For CSL, we use the same bounds on $\lambda_k$ and $\lambda_k/\sigma^2$ as for GRW, but then combine with \eqref{rateCSL}. The bounds thus obtained from different experiments are collected in Table~\ref{table:dif}.

\begin{sidewaystable}[htbp]
\label{table:dif}
{\footnotesize
\begin{tabular}{llccr@{}l@{\,}lr@{}l@{}lr@{}l@{}lr@{}l@{\,}lr@{}l@{}lr@{}l@{\,}l}
\hline\noalign{\medskip}
Year & first author [ref.] & interfering & $m/m_p$ & \multicolumn{3}{c}{$\tau$} & \multicolumn{3}{c}{$d$} & \multicolumn{3}{c}{in GRW} & \multicolumn{3}{c}{in GRW} & \multicolumn{3}{c}{in CSL} & \multicolumn{3}{c}{in CSL}\\
& & object & & & & & & & & \multicolumn{3}{c}{$\lambda<$} & \multicolumn{3}{c}{$\lambda/\sigma^2<$} & \multicolumn{3}{c}{$\lambda<$} & \multicolumn{3}{c}{$\lambda/\sigma^2<$} \\[2mm]\hline\hline\noalign{\medskip}
1927 & Davisson 
\cite{DG27} & electron & $5\times10^{-4}$ & \multicolumn{3}{c}{N/A} 
& $2\times$ & $10^{-10}$ & $\mathrm{m}$ 
& & $10^{14}$&$\mathrm{s}^{-1}$ 
& $3\times$ & $10^{33}$&$\mathrm{m}^{-2}\mathrm{s}^{-1}$ 
& &$10^{17}$ & $\mathrm{s}^{-1}$ 
& $5\times$ & $10^{36}$ & $\mathrm{m}^{-2}\mathrm{s}^{-1}$\\[2mm]
1930 & Estermann \cite{ES30} &He&4
&\multicolumn{3}{c}{N/A} 
&$4\times$&$10^{-10}$&$\mathrm{m}$ 
&&$10^{11}$&$\mathrm{s}^{-1}$  
&$6\times$&$10^{29}$&$\mathrm{m}^{-2}\mathrm{s}^{-1}$
&$3\times$&$10^{10}$&$\mathrm{s}^{-1}$
&&$10^{29}$&$\mathrm{m}^{-2}\mathrm{s}^{-1}$ \\[2mm]
1959 & M\"ollenstedt \cite{MJ59} &electron&$5\times10^{-4}$
&$3\times$&$10^{-9}$&$\mathrm{s}$ 
&$2\times$&$10^{-6}$&$\mathrm{m}$ 
&$7\times$&$10^{11}$&$\mathrm{s}^{-1}$
&&$10^{23}$&$\mathrm{m}^{-2}\mathrm{s}^{-1}$
&&$10^{15}$&$\mathrm{s}^{-1}$
&$3\times$&$10^{26}$&$\mathrm{m}^{-2}\mathrm{s}^{-1}$\\[2mm]
1987 & Tonomura \cite{TEEMK89} & electron & $5\times10^{-4}$ 
& &$10^{-8}$ & $\mathrm{s}$ 
& &$10^{-4}$&$\mathrm{m}$ 
&$2\times$ &$10^{11}$&$\mathrm{s}^{-1}$ 
&$2\times$ &$10^{19}$&$\mathrm{m}^{-2}\mathrm{s}^{-1}$ 
&$4\times$ &$10^{14}$&$\mathrm{s}^{-1}$ 
&$4\times$ &$10^{22}$&$\mathrm{m}^{-2}\mathrm{s}^{-1}$\\[2mm]
1988 & Zeilinger \cite{Zei88} & neutron & 1 
& & $10^{-2}$&$\mathrm{s}$ 
& & $10^{-4}$&$\mathrm{m}$ 
& $2\times$& $10^2$&$\mathrm{s}^{-1}$ 
&$2\times$ & $10^{10}$ & $\mathrm{m}^{-2}\mathrm{s}^{-1}$ 
& $2\times$& $10^2$ & $\mathrm{s}^{-1}$ 
& $2\times$& $10^{10}$&$\mathrm{m}^{-2}\mathrm{s}^{-1}$\\[2mm]
1991 & Carnal \cite{CM91} &He&4
&$6\times$&$10^{-4}$&$\mathrm{s}$ 
&&$10^{-5}$&$\mathrm{m}$ 
&$4\times$&$10^2$&$\mathrm{s}^{-1}$
&$4\times$&$10^{12}$&$\mathrm{m}^{-2}\mathrm{s}^{-1}$
&&$10^2$&$\mathrm{s}^{-1}$
&&$10^{12}$&$\mathrm{m}^{-2}\mathrm{s}^{-1}$\\[2mm]
1999 & Arndt \cite{Zei99} &C$_{60}$&720
&$6\times$&$10^{-3}$&$\mathrm{s}$ 
&&$10^{-7}$&$\mathrm{m}$ 
&$2\times$&$10^{-1}$&$\mathrm{s}^{-1}$ 
&$2\times$&$10^{13}$&$\mathrm{m}^{-2}\mathrm{s}^{-1}$
&$3\times$&$10^{-4}$&$\mathrm{s}^{-1}$
&$3\times$&$10^{10}$&$\mathrm{m}^{-2}\mathrm{s}^{-1}$ \\[2mm]
2001 & Nairz \cite{Zei01} &C$_{70}$&840
&&$10^{-2}$&$\mathrm{s}$ 
&$3\times$&$10^{-7}$&$\mathrm{m}$
&&$10^{-1}$&$\mathrm{s}^{-1}$
&&$10^{12}$&$\mathrm{m}^{-2}\mathrm{s}^{-1}$
&&$10^{-4}$&$\mathrm{s}^{-1}$
&&$10^9$&$\mathrm{m}^{-2}\mathrm{s}^{-1}$ \\[2mm]
2004 & Hackerm\"uller \cite{Ha04} & C$_{70}$ & 840 
& $2\times$ & $10^{-3}$&$\mathrm{s}$ 
& & $10^{-6}$&$\mathrm{m}$ 
& & $10^0$ &$\mathrm{s}^{-1}$ 
& & $10^{12}$&$\mathrm{m}^{-2}\mathrm{s}^{-1}$ 
& & $10^{-3}$&$\mathrm{s}^{-1}$ 
&& $10^9$&$\mathrm{m}^{-2}\mathrm{s}^{-1}$ \\[2mm]
2007 & Gerlich \cite{Ge07} & C$_{30}$H$_{12}$F$_{30}$N$_2$O$_4$ 
&$10^3$ 
&&$10^{-3}$&$\mathrm{s}$ 
&$3\times$&$10^{-7}$&$\mathrm{m}$ 
&&$10^{0}$&$\mathrm{s}^{-1}$ 
&&$10^{13}$&$\mathrm{m}^{-2}\mathrm{s}^{-1}$ 
&&$10^{-3}$&$\mathrm{s}^{-1}$ 
&&$10^{10}$&$\mathrm{m}^{-2}\mathrm{s}^{-1}$\\[2mm]
2011 & Gerlich \cite{Ge11} & C$_{60}[$C$_{12}$F$_{25}]_{10}$ &
$7\times 10^3$ 
&&$10^{-3}$&$\mathrm{s}$ 
&$3\times$&$10^{-7}$&$\mathrm{m}$ 
&&$10^{-1}$&$\mathrm{s}^{-1}$
&&$10^{12}$&$\mathrm{m}^{-2}\mathrm{s}^{-1}$
&&$10^{-5}$&$\mathrm{s}^{-1}$
&&$10^8$&$\mathrm{m}^{-2}\mathrm{s}^{-1}$ \\[2mm]
\hline\noalign{\medskip}
&\multicolumn{21}{l}{Proposed future experiments} \\[2mm]
\hline\hline\noalign{\medskip}
&Romero-Isart \cite{Rom11} & [SiO$_2$]$_{150,000}$&
$10^7$ 
&&$10^{-1}$&$\mathrm{s}$ 
&$4\times$&$10^{-7}$&$\mathrm{m}$
&&$10^{-6}$&$\mathrm{s}^{-1}$
&$6\times$&$10^{6}$&$\mathrm{m}^{-2}\mathrm{s}^{-1}$
&&$10^{-13}$&$\mathrm{s}^{-1}$
&$6\times$&$10^{-1}$&$\mathrm{m}^{-2}\mathrm{s}^{-1}$ \\[2mm]
&Nimmrichter \cite{Ni11} & Au$_{500,000}$ &
$10^8$ 
&$6\times$&$10^{0}$&$\mathrm{s}$ 
&&$10^{-7}$&$\mathrm{m}$ 
&$2\times$&$10^{-9}$&$\mathrm{s}^{-1}$
&$2\times$&$10^{5}$&$\mathrm{m}^{-2}\mathrm{s}^{-1}$
&$2\times$&$10^{-17}$&$\mathrm{s}^{-1}$
&$2\times$&$10^{-3}$&$\mathrm{m}^{-2}\mathrm{s}^{-1}$ \\[2mm]
\hline
\end{tabular}
}
\caption{Bounds on $\sigma,\lambda$ obtained from different diffraction experiments. For each experiment, $m$ = mass of the interfering object, $m_p$ = proton mass, $\tau$ = time of flight between grating and image plane, $d$ = period of grating (or transverse coherence length in \cite{TEEMK89}), N/A = not applicable. For each theory (GRW or CSL), two bounds are obtained. This table is the basis for Fig.~\ref{fig:3}.}
\end{sidewaystable}

\subsection{Universal Warming}

Because the collapses tend to add energy to every system, temperatures of all things increase. 
A basic formula specifies the average energy injected by a GRW collapse hitting a free particle of mass $m$ \cite{GRW86}:
\be\label{Ecollapse}
\Delta E = \frac{3\hbar^2}{4m} \frac{1}{\sigma^2}\,.
\ee
Thus, the energy of a system of $N$ free particles of mass $m$ tends to increase in the average in a GRW world at the rate
\be\label{GRWErate}
\frac{dE}{dt} = \frac{3\hbar^2}{4} \frac{N}{m_p} \frac{\lambda}{\sigma^2}\,,
\ee
with an additional factor $m/m_p$ in CSL.
By virtue of the relation $E=\tfrac{3}{2}k_BNT$,  the temperature of a gas of free particles of mass $m$ increases at the rate
\be\label{tempincrease}
\frac{dT}{dt} = \frac{\hbar^2}{8k_B m_p}\frac{\lambda}{\sigma^2}\,,
\ee
with an additional factor $m/m_p$ in CSL.

Upper bounds on the actual rate of temperature increase could be obtained in many ways from observations of absence of warming of different things, e.g., a cup of cold water, the ocean, the atmosphere, other planets, or the intergalactic medium. The biggest problem with obtaining good bounds is to control the possible ways of cooling. 
\begin{itemize}
\item A liter of ice cold water, isolated from an environment at room temperature by means of walls of styrofoam of thickness (say) $5\,\mathrm{cm}$ equilibrates with the environment in about one day in the absence of spontaneous warming. Spontaneous warming will not speed up the observable temperature increase drastically provided its rate is less than $10\, \mathrm{K}/\mathrm{day}$.

\item We are not confident about controlling how much energy the ocean, the atmosphere, or other planets radiate off into space, nor about the temperature balance of the intergalactic medium (IGM). Adler \cite{Adl07} has assessed the cooling mechanisms of the IGM at a distance from Earth corresponding to a red shift of $z=3$. According to his analysis, the dominant mechanism of cooling is adiabatic expansion and amounts to an energy reduction of $5\times 10^{-17} \, \mathrm{eVs}^{-1} = 8 \times 10^{-36}\,\mathrm{Js}^{-1}$ per proton (the IGM consists of highly ionized hydrogen). Assuming that the temperature of the IGM remains constant (at the observed value of about $2\times 10^4\, \mathrm{K}$), the above energy loss rate represents an upper bound on the energy gain rate \eqref{GRWErate} with $N=1$ and $m=m_p$, implying
\be
\frac{\lambda}{\sigma^2} < 2\times 10^6 \, \mathrm{m}^{-2}\mathrm{s}^{-1}
\ee
for both GRW and CSL. We remain cautious about this bound because its validity depends on the correct assessment of the relevant mechanisms of cooling. That is why we have not included it in Fig.~\ref{fig:1}, though we have drawn it as a dashed line in Fig.~\ref{fig:2}. 

\item Here is a bound that does not require controlling the mechanisms of cooling and thus is more certain. The temperature in the biggest natural cave of Germany, the Kubacher Kristallh\"ohle, is $9^{\circ} \,\mathrm{C}$ all year around \cite{Kub}, and thus is sometimes below the temperature at the surface. During July, the lowest surface temperatures in Germany usually stay above $13^\circ\,\mathrm{C}$ \cite{climate}. Thus, heat spontaneously created in the cave cannot be transported away: neither to the surface because it is warmer nor to the kernel of the Earth because it is hot. Given that the temperature in the cave does not increase by more than $1\, \mathrm{K}$ during July, we obtain the empirical bound
\be\label{tempbound}
\frac{dT}{dt} < 3\times 10^{-2}\,\mathrm{K}
\ee
on the rate of spontaneous warming.

Using \eqref{tempincrease} for air with $m=28m_p$ the mass of an $N_2$ molecule, \eqref{tempbound} yields the following bound:
\be
\frac{\lambda}{\sigma^2} < 
\begin{cases}
10^{13}  \, \mathrm{m}^{-2}\mathrm{s}^{-1} & \text{GRW}\\
3\times 10^{11} \, \mathrm{m}^{-2}\mathrm{s}^{-1} & \text{CSL.}
\end{cases}
\ee
More or less the same bound is obtained for water or rock instead of air if we assume that the rate of temperature increase in water or rock is comparable to \eqref{tempincrease} with $m=18m_p$ (the mass of an H$_2$O molecule) or $m=28m_p$ (the mass of a silicon atom).
\end{itemize}

\subsection{Spontaneous X-Ray Emission}

Further empirical bounds arise from an experiment in which the rate of spontaneous 11 keV photon emission from germanium, as monitored in 1 keV bins, has been bounded by 0.05 pulses/(keV kg day). According to the analyses by Collett et al.~\cite{CPAN95,CP03}, Fu \cite{Fu}, and Adler \cite{Adl07}, this requires
\be
\frac{\lambda}{\sigma^4} < 10^{26} \,\mathrm{m}^{-4}\mathrm{s}^{-1} 
\text{ and }
\frac{\lambda}{\sigma^2} < 10^{7} \, \mathrm{m}^{-2}\mathrm{s}^{-1}
\text{ in GRW,}
\ee
\be
\frac{\lambda}{\sigma^4} < 10^{26} \,\mathrm{m}^{-4}\mathrm{s}^{-1}
\text{ and }
\frac{\lambda}{\sigma^2} < 10^{10} \, \mathrm{m}^{-2}\mathrm{s}^{-1}
\text{ in CSL.}
\ee

\subsection{Spontaneous Sound Emission}

For sufficiently small $\sigma$, every single GRW collapse would inject so much energy into the particle affected that a noticeable explosion would occur, which should lead to the emission of sound (besides radiation and heat). The fact that we do not hear spontaneous bangs leads to bounds on $\sigma$ and $\lambda$ as follows. One can hear a bang of energy $10^{-6}\, \mathrm{J}$ (which corresponds to the click of a typewriter \cite{sound}) or more. If we assume that the energy injected by collapse into an electron bound in an atom is comparable to that for a free electron as in Eq.~\eqref{Ecollapse}, and that a substantial fraction of it is emitted as sound, then we obtain that a single collapse will cause an audible noise for $\sigma<10^{-16}\,\mathrm{m}$. We assume that spontaneous bangs would have been noticed if they occurred more than once per month within $5\,\mathrm{m}$ distance, that is, if the collapse rate in the volume $(2\pi/3)(5\,\mathrm{m})^3$ (or $10^4$ moles) of air, which is (number of moles) $\times$ (number of molecules per mole) $\times$ (number of electrons per N$_2$ molecule) $\times$ (collapse rate per electron) = $10^4 \times 6 \times 10^{23} \times 14 \times 1800^{-1}\, \lambda = 4.7 \times 10^{25}\lambda$, exceeded $4\times 10^{-7} \,\mathrm{s}^{-1}$. Thus, bangs would have been noticed if
\be
\sigma<10^{-16}\,\mathrm{m}\mbox{ and }\lambda > 10^{-32}\,\mathrm{s}^{-1}\,.
\ee
This region is covered by the region refuted by x-ray experiments, as well as by the region refuted by the absence of air heating (together with the PUR), and for this reason is not drawn in the figures.

\subsection{Decay of Supercurrents}

Rae \cite{Rae90} and others \cite{BNR95,Leg02,Adl07} have pointed to empirical consequences of GRW/CSL concerning supercurrents in a superconducting ring. Since spontaneous collapses would break Cooper pairs, the supercurrent would spontaneously decay (unless the Cooper pairs get re-created) at a rate, according to the analyses of Rae \cite{Rae90}, Buffa, Nicrosini, and Rimini \cite{BNR95}, and Adler \cite{Adl07}, of
\be
\frac{1}{\sigma k_F} \frac{m_e}{m_p}  \lambda
\ee
(up to a factor of $3/2\sqrt{\pi}=0.85$ \cite[Eq.~(5.11)]{BNR95}) in GRW, with an additional factor $m_e/m_p$ in CSL. Here, $m_e$ is the electron mass and $\hbar k_F$ the Fermi momentum, with $k_F= 1.6\times 10^{10}\,\mathrm{m}^{-1}$ for a realistic setup \cite[Eq.~(3.11)]{BNR95}.  The factor $1/\sigma k_F$ arises from the indistinguishability of the electrons and thus applies in the version of GRW that we are using as well as in CSL.

Experiments suggest that the supercurrent actually decays no faster than at a rate of $3\times 10^{-13}\,\mathrm{s}^{-1}$ \cite{Rae90,Adl07}. This result would imply the bound
\begin{align}
\label{squidbound} \frac{\lambda}{\sigma}& < 
\begin{cases} 10\, \mathrm{m}^{-1}\mathrm{s}^{-1} &\text{GRW}\\ 2\times 10^4\,\mathrm{m}^{-1}\mathrm{s}^{-1}&\text{CSL} \end{cases}
\text{ for }\sigma\leq10^{-3}\,\mathrm{m},\\
\label{squidbound2} \frac{\lambda}{\sigma^3} &< 
\begin{cases} 10^{7}\, \mathrm{m}^{-3}\mathrm{s}^{-1} &\text{GRW}\\ 2\times 10^{10}\,\mathrm{m}^{-3}\mathrm{s}^{-1}&\text{CSL} \end{cases}
\text{ for }\sigma>10^{-3}\,\mathrm{m}.
\end{align}
However, this bound may be too low because it does not take the possible re-creation of Cooper pairs into account. For this reason we have not included it in Fig.s \ref{fig:1} and \ref{fig:GRWf}; we have drawn it in Fig.~\ref{fig:2} using dashed-and-dotted lines.

\section{Considerations for Drawing the PUR}
\label{sec:PUR}

We regard a parameter choice $(\sigma,\lambda)$ as philosophically satisfactory if and only if the PO agrees on the macroscopic scale with what humans normally think macroscopic reality is like. This criterion differs from the following others which have been or might have been suggested, and which we find less convincing: (i)~Systems that are commonly regarded as ``classical'' practically do not occur in superpositions of their classical states. (ii)~States of systems that are commonly regarded as measurement outcomes practically do not occur in superpositions of states corresponding to different outcomes. (iii)~Superpositions of different perceptions of human beings practically do not occur. Adler's \cite{Adl07} and Gisin and Percival's \cite{GP93} view of what is philosophically satisfactory, elucidated in terms of latent image formation, seems linked to (i) and/or (ii), as a latent image is ``a permanent classical record of a quantum event'' \cite{GP93}. 

Our criterion still requires a decision as to where the ``macroscopic scale'' begins. On the practical side, there should not be much disagreement. If decisions made by other authors differ by factors of 10 or even 100, then it will still not be a dramatic change in the diagram, as evident from Fig.~\ref{fig:1} and Fig.~\ref{fig:GRWf}. In fact, such differences reflect the natural fuzziness of the concept of the ``macroscopic.'' On the principled side, however, Bassi and Pearle (personal communication) have raised the question whether the choice of macroscopic scale should depend on the physical properties of human beings (e.g., on their reaction time and the resolution of their eyes). We tend to think it should, even though our criterion is different from (iii).

We now estimate the boundary fo the PUR explicitly, first for GRWm, i.e., the version of GRW with the PO given by the matter density function $m(x,t)$ (as in Fig.~\ref{fig:1}). If the collapse rate $\lambda$ is sufficiently small then a superposition $\psi=\sum_i c_i \psi_i$ of macroscopically different contributions $\psi_i$ (such as Schr\"odinger's cat) will fail to quickly decay to one of the $\psi_i$. In that case, all of the $\psi_i$ contribute to the $m$ function, giving the theory a many-worlds character. Indeed, in the limit $\lambda\to0$ the theory approaches ``Sm,'' Schr\"odinger's many-worlds theory \cite{AGTZ11}. If Sm is regarded as satisfactory then the PUR should in a sense be regarded as empty, since GRWm/CSLm then remains satisfactory even for very small $\lambda$. However, the very-small-$\lambda$ regime defies the purpose that collapse theories were introduced for, and it would then be simpler and more natural to set $\lambda=0$ and adopt Sm. That is why we include such points in PUR (and thus draw it as non-empty), although we do not take a position here as to whether Sm is satisfactory. 

Let us formulate more explicitly which points we include in the PUR. Note that collapse theories and many-worlds theories are intended to use different ways out of the measurement problem of quantum mechanics: In collapse theories, superpositions of different outcomes do not arise (for all practical purposes), whereas in many-worlds theories there is no unique actual outcome. We include those points in the PUR for which the former solution to the measurement problem is not realized, i.e., those for which macroscopic superpositions are not avoided. 

This situation occurs in GRWm either if too few collapses happen (i.e., if $\lambda$ is too small) or if the collapses are too mild compared to their frequency (i.e., if $\sigma$ is too big, given $\lambda$) because a very mild (i.e., large-$\sigma$) collapse has a very weak effect on the wave function. By \eqref{lambdasigma2}, the latter case occurs if $\lambda/\sigma^2$ is too small.

To obtain quantitative estimates for the values of $\lambda$ and $\lambda/\sigma^2$ that define the boundary of the PUR, we ask under which conditions measurement outcomes can be read off unambiguously from the $m$ function. For definiteness, we think of the outcome as a number printed on a sheet of paper; we estimate that a single digit, printed (say) in 11-point font size, consists of $3\times10^{17}$ carbon atoms or $N=4\times 10^{18}$ nucleons.\footnote{Here is how this estimate was obtained: We counted that a typical page (from the Physical Review) without figures or formulas contains 6,000 characters and measured that a toner cartridge for a Hewlett Packard laser printer weighs 2.34 kg when full and 1.54 kg when empty. According to the manufacturer, a cartridge suffices for printing $2\times 10^4$ pages. Assuming that the toner consists predominantly of carbon, we arrive at $3\times 10^{17}$ atoms per character.} 
The rate of collapse on a body consisting of $N$ nucleons is
\be
\Gamma=N\lambda\,.
\ee
For definiteness, we choose the maximal time for which macroscopic superpositions can be tolerated to be half a second. To ensure that a superposition of (say) ``2'' and ``3'' decays in that time, a superposition involving position differences of $10^{-3}\,\mathrm{m}$, we need that
\be\label{Gammalimits}
\Gamma>2\,\mathrm{s}^{-1}\mbox{ and }
\frac{\Gamma}{\sigma^2}>2\times 10^{6}\,\mathrm{m}^{-2}\mathrm{s}^{-1}
\ee
or
\be
\lambda>5\times10^{-19}\,\mathrm{s}^{-1}\mbox{ and }
\frac{\lambda}{\sigma^2}>5\times 10^{-13}\,\mathrm{m}^{-2}\mathrm{s}^{-1}\,. 
\ee
This is how we obtain the boundary of the PUR for GRWm.

\medskip

For CSLm, the collapse rate for an object consisting of $N$ nucleons, of which groups of $n$ are closer than $\sigma$ is \cite{Adl07}
\be
\Gamma = nN\lambda \,.
\ee
In our consideration of a single character, we have that $N=4\times 10^{18}$ and require \eqref{Gammalimits}. Assuming that the ink layer of a printed character is about $10^{-5}\,\mathrm{m}$ thick, that a line in a character is about $10^{-4}\,\mathrm{m}$ wide and a few millimeters long, and that the ink contains about $10^{30}$ nucleons per $\mathrm{m}^3$ of volume, we conclude that the number $n$ of nucleons belonging to the ink within distance $\sigma$ is
\be
n=\begin{cases}
\frac{4\pi}{3}\sigma^3 \times 10^{30} \, \mathrm{m}^{-3} 
	& \text{if }\sigma < 10^{-5}\,\mathrm{m}\\
\pi\sigma^2 \times 10^{25}\,\mathrm{m}^{-2} 
	& \text{if }10^{-5}\,\mathrm{m}<\sigma< 10^{-4}\,\mathrm{m}\\
\sigma \times 4\times 10^{21} \, \mathrm{m}^{-1}
	& \text{if } 10^{-4}\,\mathrm{m}<\sigma<10^{-3}\,\mathrm{m}\\
4\times 10^{18} 
	& \text{if } 10^{-3}\,\mathrm{m}<\sigma\,.
\end{cases}
\ee
We thus obtain
\be
\lambda> \begin{cases}
\sigma^{-3}\times 10^{-49}\,\mathrm{s}^{-1}\mathrm{m}^3 
	&\text{if }\sigma<10^{-5}\,\mathrm{m}\\ 
\sigma^{-2}\times 10^{-44}\,\mathrm{s}^{-1}\mathrm{m}^2
	& \text{if }10^{-5}\,\mathrm{m}<\sigma<10^{-4}\,\mathrm{m}\\
\sigma^{-1} \times 10^{-40}\,\mathrm{s}^{-1}\mathrm{m}^{-1}
	& \text{if }10^{-4}\,\mathrm{m}<\sigma<10^{-3}\,\mathrm{m}\\
\sigma^2 \times 10^{-31}\,\mathrm{s}^{-1}\mathrm{m}^{-2}
	& \text{if }10^{-3}\,\mathrm{m}<\sigma
\end{cases}
\ee
defining the boundary of the PUR in CSLm.

\medskip

Let us now consider GRWf, i.e., the GRW theory with the PO given by flashes. Macroscopic objects (say, chairs) in 3-dimensional space are to be found in the pattern of flashes. Since one flash occurs at every collapse, very small values of $\lambda$ mean that the flashes per second are too few to contain a chair, which makes the theory philosophically unsatisfactory.\footnote{\label{fn:huge}One could argue that the theory actually becomes empirically refuted, as it predicts the non-existence of chairs while we are sure that chairs exist in our world. However, this empirical refutation can never be conclusively demonstrated because the theory would still make reasonable predictions for the outcomes of all experiments, provided the outcomes are displayed using sufficiently many particles (say, using huge letters made of rock; ``sufficiently many'' means more than roughly $1\,\mathrm{s}^{-1}/\lambda$). That is why we need to distinguish between the kind of empirical inadequacy that can be demonstrated (on which the ERR is based) and the kind just described (which can be discovered only theoretically and which we include in the PUR).} So for GRWf, $\lambda\to 0$ does not mean many-worlds but zero-worlds. To make a quantitative judgement about the critical value of $\lambda$, we consider again a digit printed on a sheet of paper over a time interval of half a second; if the number of flashes is less than 10, its value is not sufficiently well defined. This happens for $\lambda<5\times10^{-18}\,\mathrm{s}^{-1}$.

GRWf is also unsatisfactory if $\sigma > 10^{-3}\,\mathrm{m}$, as the locations of the flashes are randomized over a length scale of $\sigma$.\footnote{Recall that the probability distribution \eqref{Zdef} of the collapse center is not $|\psi|^2$ itself, but $|\psi|^2$ smeared out with a Gaussian of width $\sigma$. By the way, the law for the $m$ function in GRWm might or might not involve such a smearing (both versions are possible). We assume in this paper that it does not, see \eqref{mdef}. For the version that does, i.e., with $\delta^3$ in \eqref{mdef} replaced by $g$, the PUR is larger and includes all points with $\sigma>10^{-3}\,\mathrm{m}$.} In this case, all detail finer than $10^{-3}\,\mathrm{m}$ would be washed out in the PO, and again the value of a digit printed on a sheet of paper would not be well defined in the PO. The PUR defined by the disjunction of the conditions $\lambda<5\times 10^{-18}\,\mathrm{s}^{-1}$ and $\sigma>10^{-3}\,\mathrm{m}$ already contains the PUR of GRWm as a subset, so we need not repeat the previous considerations for GRWm.\footnote{The fact that the PUR of GRWf is different from that of GRWm illustrates that they are different theories, contrary to a suggestion by D. Z. Albert. Readers may also wonder how the PURs can differ, given that GRWf and GRWm have been proved to be empirically equivalent \cite{AGTZ06,grw3A}: The answer is that the proof shows that for any experiment, GRWf and GRWm \emph{ultimately} agree, i.e., they agree provided the outcomes are displayed for a sufficiently long time using sufficiently many particles, as described in Footnote~\ref{fn:huge}.}

What if future experiments falsified quantum mechanics and confirmed GRW or CSL for a particular parameter pair that lies in the PUR? For GRWf we would conclude that the chosen PO is wrong. For GRWm/CSLm, either the PO is wrong or reality has a many-worlds character.

\bigskip

\noindent\textit{Acknowledgments.} 
We thank Joshua Smith for practical assistance, Anthony Leggett and Tim Maudlin for helpful discussions, and Angelo Bassi and Philip Pearle for references and useful comments on a previous version of this article. 
W.F. acknowledges support from DIMACS at Rutgers, the State University of New Jersey. R.T. acknowledges support from NSF Grant SES-0957568 and the Trustees Research Fellowship Program at Rutgers. 

\end{document}